\documentclass[twocolumn,showpacs,aps,prl,floatfix]{revtex4}
\usepackage{graphicx}
\usepackage{subfigure}
\usepackage{epsfig}
\usepackage{psfrag}

\makeatletter

\usepackage{verbatim}

\makeatletter

\input epsf

\def\be{\begin{equation}}
\def\ee{\end{equation}}
\def\ba{\begin{eqnarray}}
\def\ea{\end{eqnarray}}

\makeatother

\begin{document}

\title{Spin-wave dispersion in half-doped  La$_{3/2}$Sr$_{1/2}$ NiO$_4$}                 

\author{D.~X.~Yao$^1$ and E.~W.~Carlson$^2$}
\affiliation{(1) Dept. of Physics, Boston University, Boston, MA 02215\\
(2) Dept. of Physics, Purdue University, West Lafayette, IN  47907}

\date{September 12, 2006}

\begin{abstract}
Recent neutron scattering measurements 
reveal spin and charge ordering in the half-doped nickelate, 
La$_{3/2}$ Sr$_{1/2}$ NiO$_4$.  Many of the features of the
magnetic excitations have been explained in terms of the 
spin waves of diagonal stripes with weak single-ion
anisotropy.  However, an  optical mode dispersing away from the $(\pi,\pi)$
point was not captured by this theory.   
We show here that this apparent optical mode is a natural consequence
of stripe twinning in a diagonal stripe pattern with a magnetic coupling structure which
is two-fold symmetric, {\em i.e.} one possessing the same spatial rotational symmetry
as the ground state.  

\end{abstract}
\maketitle

Strongly correlated electronic systems often exhibit some evidence of charge, 
spin, or orbital order, or some combination thereof.  
The interplay between these degrees of freedom can lead
to a wide variety of novel phases.  
The nickelates in particular show a wide range of doping
in which both charge and spin order coexist in stripe patterns.
The less-ordered stripe structures in some families of 
cuprates have been widely studied for their possible connection
to high temperature superconductivity. 

Recent experiments on
doped La$_{2-x}$ Sr$_{x}$ NiO$_4$ have shown clear evidence of static
diagonal charge and spin stripe order.\cite{boothroyd03a, boothroyd05} 
Spin wave theory has been successful at describing much of the 
behavior in these spin-ordered systems.\cite{erica04,yaocarlson06b,tranquada,tranquada03,kruger03,freeman05a} 
We consider here the recent experiments by Freeman {\em et al.} on
the spin dynamics of 
half-doped La$_{3/2}$ Sr$_{1/2}$NiO$_4$  using inelastic
neutron scattering.\cite{freeman05a} 
In this material, the spins are in a diagonal stripe phase,
where stripes run $45^o$ from the Ni-O bond direction, 
and the charged domain walls are only 2 lattice constants apart.
The charge density modulation  can either be considered as densely packed stripes, or as  
a checkerboard, for domain walls centered on the Ni sites,
since in that case the two ways of describing the charge pattern
are indistinguishable at this filling.  
However, if the domain walls are centered on oxygen sites ({\em i.e.} for bond-centered stripes), 
the charge checkerboard description is not applicable.  

In this paper, we are interested in the extra magnetic mode
dispersing away from the antiferromagnetic wavevector $Q_{\rm AF} = (0.5,0.5)$
above $50$meV in Fig.~3 of Ref.~\onlinecite{freeman05a}.
One explanation put forth by the authors is that diagonal discommensurations
in the spin order may be able to account for this extra scattering mode.  
We show here that the mode could also be due to asymmetry in the 
spin coupling constants, in which case the ``extra mode'' is really an
extension of the acoustic band, made visible due to stripe twinning.

\begin{figure}[htb]
{\centering
  \subfigure[ DS2]
  {\resizebox*{0.45
 \columnwidth}{!}{\includegraphics{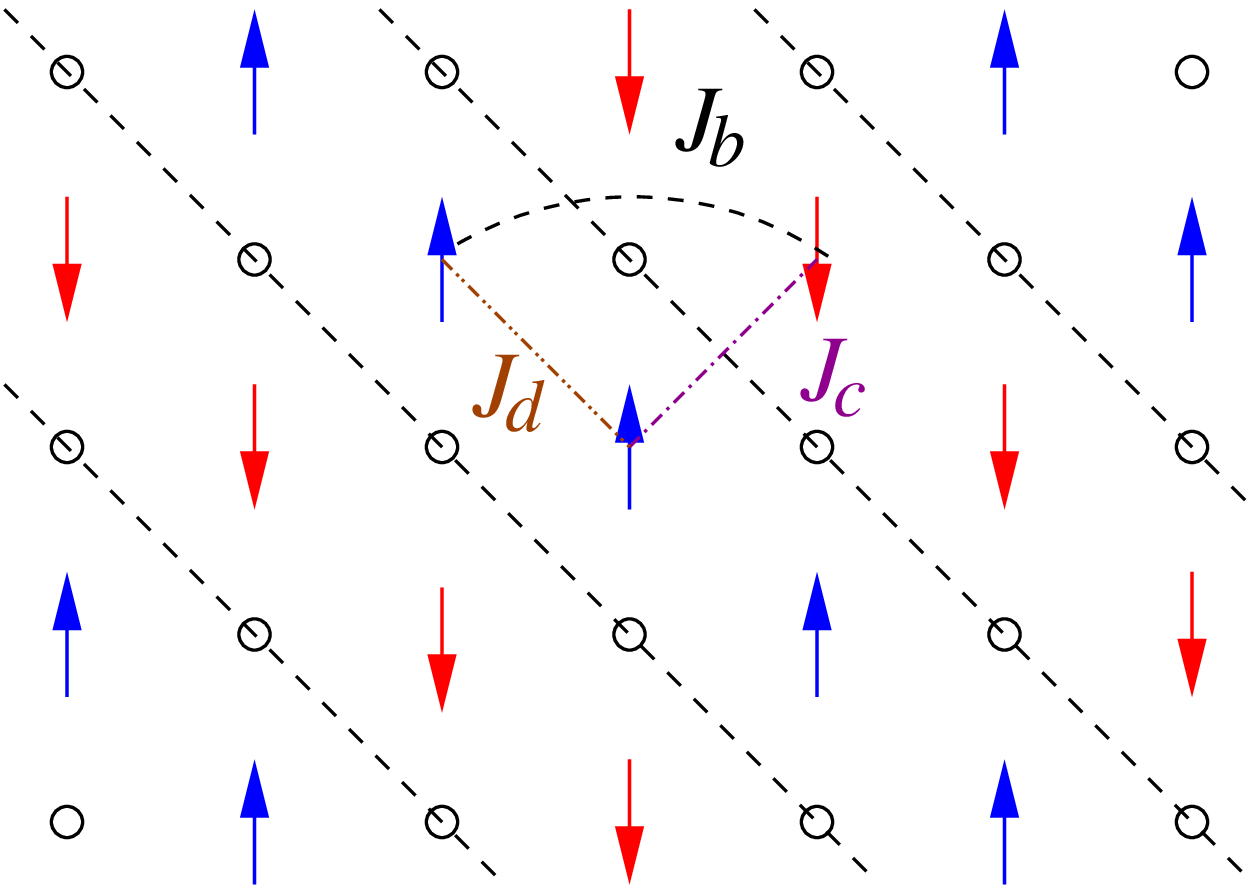}\label{ds2.lattice}}}
 \hspace{.2in}
  \subfigure[ DB2]
  {\resizebox*{0.4\columnwidth}{!}{\includegraphics{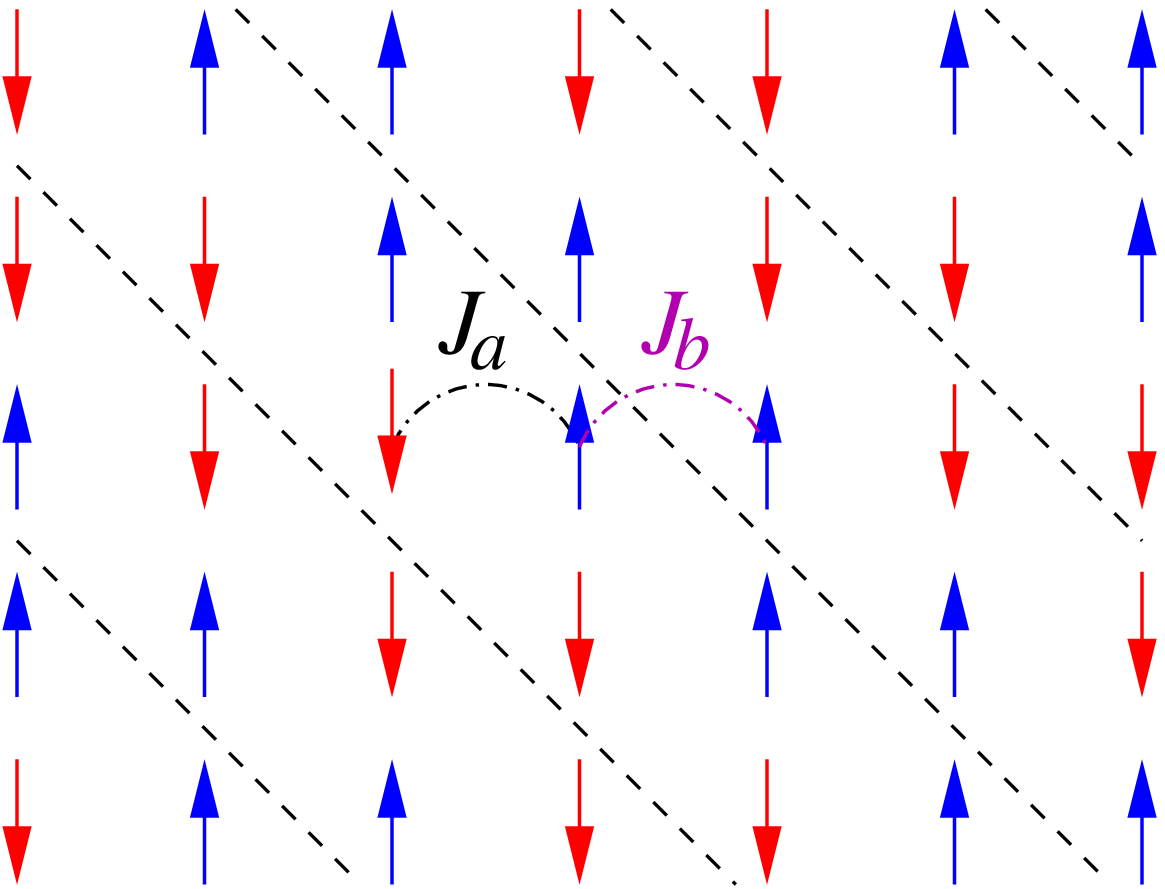}\label{db2.lattice}}}
  \par}
\caption{(Color online) Spin-charge ordering in a NiO$_2$ square
  lattice. Arrows represent spins on Ni$^{2+}$ sites and circles represent
  Ni$^{3+}$ holes. 
  (a) DS2:  Diagonal, site-centered stripes of spacing $p=2$.  
  Note that in this configuration the charges distribution is 
symetric under  $90^o$ rotations, and is equivalent to a  checkerboard pattern.
  The spin configuration is symmetric under $180^o$ rotations, and breaks the symmetry of the charge checkerboard.  The straight-line exchange coupling  across the charge domain wall is $J_b$, 
  and   $J_c$ and $J_d$ are diagonal couplings perpendicular and
  parallel to the charge domain wall, respectively.  
(b) DB2:  Diagonal, bond-centered stripes of spacing $p=2$.  
Here, both the charge and spin patterns are 
$2$-fold rotationally symmetric.
\label{lattice}}
\end{figure}

The observed ordering vector is $Q=(0.275,0.275)$, which is
close to a commensurate stripe value of $Q=(0.25,0.25)$.
The slight deviation from commensurability is believed to be due
to the discommensurations described above.  Since we are interested
in describing relatively high energy effects, in what follows
we neglect the small incommensurability, and consider commensurate diagonal stripe
structures of spacing $p=2$.  
We consider two patterns in this paper: site-centered stripes as shown in
Fig.~\ref{lattice}(a), and the corresponding bond-centered stripes shown in 
Fig.~\ref{lattice}(b).  

To model these two systems, we use a suitably parametrized 
Heisenberg model 
on a square lattice,
\begin{equation}
H=   \frac{1}{2}\sum_{i,j} J_{i,j} \mathbf{S}_{i} \cdot \mathbf{S}_{j}
\label{model}
\end{equation}
where the indices $i$ and $j$ run over all sites, and 
the couplings $J_{i,j}$ are illustrated in 
Fig.~\ref{lattice}.
For diagonal, site-centered stripes of spacing $p=2$ (DS2),
there is no need for nearest-neighbor coupling, and so we set $J_a = 0$.
The straight-line coupling $J_b$ across the domain wall is antiferromagnetic.
The diagonal coupling $J_c$ across the domain wall is also antiferromagnetic,
but the diagonal coupling $J_d$ parallel to the stripes we take to be ferromagnetic, 
$J_d <0$, as explained below.  In the diagonal bond-centered case (DB2), the nearest neighbor coupling $J_a >0$
is finite and antiferromagnetic.  We also include the ferromagnetic coupling $J_b <0$
across the domain wall.  
Since we are interested in describing high energy effects, we neglect 
the very weak single-ion anisotropy term, which splits the $2$-fold degenerate acoustic band
at low energy, with one mode remaining gapless at the IC point $Q_{\rm IC} = (0.25,0.25)$,
and the other mode developing a small gap.\cite{freeman05a}

\begin{figure}[tb]
\psfrag{6}{\Large$6$}
\psfrag{4}{\Large$4$}
\psfrag{2}{\Large$2$}
\psfrag{0.25}{\Large$0.25$}
\psfrag{0.5}{\Large$0.5$}
\psfrag{0.75}{\Large$0.75$}
\psfrag{1}{\Large$1$}
\psfrag{0}{\Large$0$}
\psfrag{w}{\LARGE$\omega$}
\psfrag{kx}{\LARGE$\hspace{-.5in}{k_x / 2\pi}$}
\psfrag{ky}{\LARGE$\hspace{.15in} {k_y / 2\pi}$}
{\centering
  \subfigure[]
  {\resizebox*{0.48\columnwidth}{!}{\includegraphics{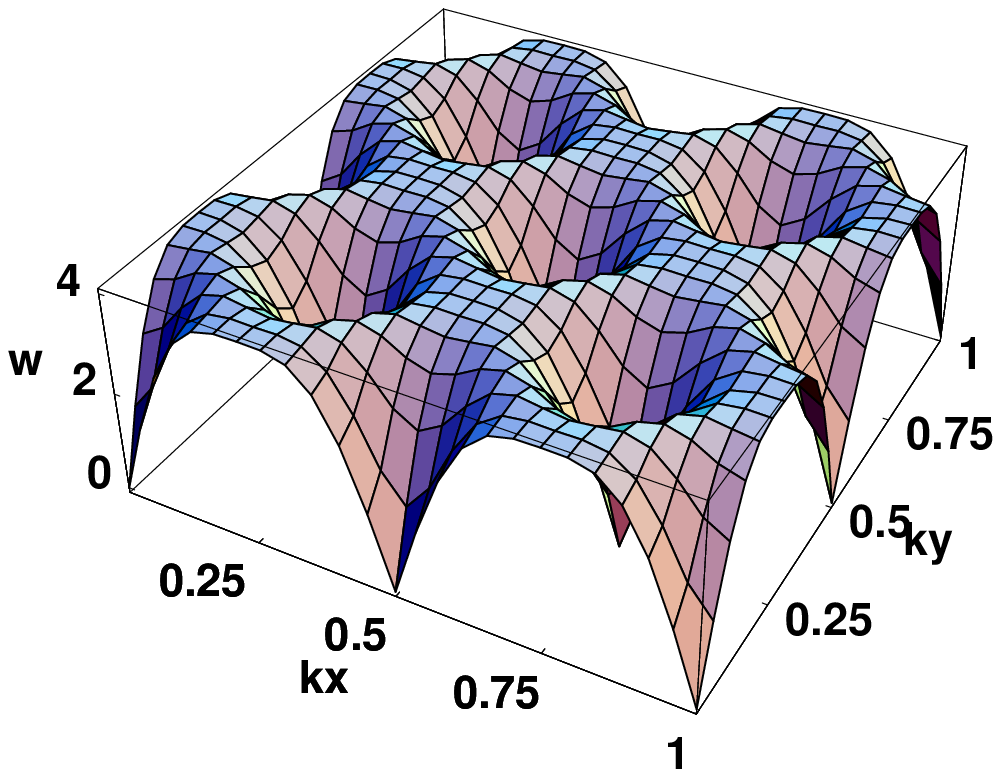}\label{af}}}
  \subfigure[]
  {\resizebox*{0.48\columnwidth}{!}{\includegraphics{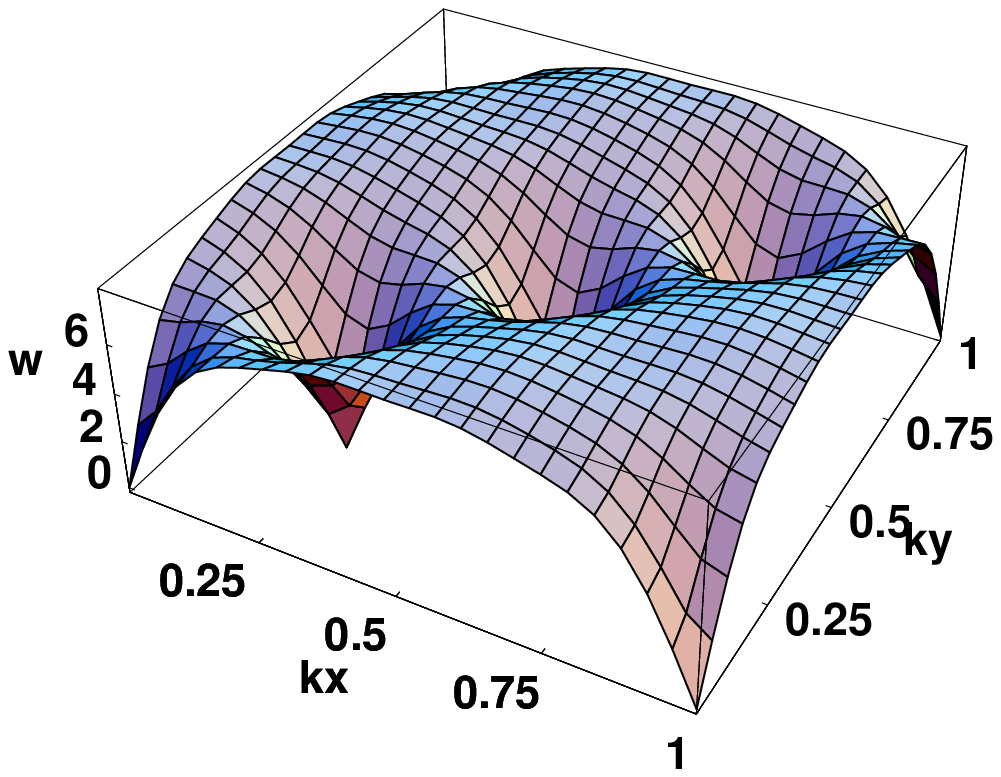}\label{uniso}}} 
  \par}
\caption{Band structures of diagonal site-centered stripes (DS2).
(a) $J_c=J_d=0$;
(b) $J_c=J_b$, $J_d=-0.5J_b$.
\label{plot3d}}
\end{figure}


For the case of site-centered stripes 
in the absence of the diagonal couplings, {\em i.e.} for $J_c=J_d=0$,
the spin system reduces to two interpenetrating antiferromagnets,
with two separate N\'eel vectors but identical N\'eel ordering temperatures.  Any weak
diagonal coupling is sufficient to establish a unique relative direction between
the two N\'eel vectors, and the ground state becomes the 
stripe structure shown in Fig.~\ref{lattice}(a).   
The number of reciprocal lattice vectors is also decreased by a factor of $2$
in the presence of the diagonal couplings $J_c \ne 0$ or $J_d \ne 0$,
as can be seen in the bandstructure of Fig.~\ref{plot3d}. 
In either case, independent of the value of $J_c$ and $J_d$,
although the antiferromagnetic point $Q_{\rm AF}$ is a magnetic reciprocal lattice vector
and therefore must have a spin wave cone dispersing out of it, there is no
net antiferromagnetism in the system, so that weight is forbidden at zero frequency
at $Q_{\rm AF}$.   The cone that emanates out of $Q_{\rm AF}$ gains finite weight
as energy is increased, but remains faint at low energies.

\begin{figure}[tb]
\psfrag{w}{$\omega$}
\psfrag{kx}{$k_x/2\pi$}
\psfrag{0.25}{$0.25$}
\psfrag{0.5}{$0.5$}
\psfrag{0.75}{$0.75$}
\psfrag{0}{$0$}
\psfrag{1}{$1$}
\psfrag{2}{$2$}
\psfrag{3}{$3$}
\psfrag{4}{$4$}
\psfrag{6}{$6$}
\psfrag{8}{$8$}
\psfrag{10}{$10$}
\psfrag{a}{(a)}
\psfrag{b}{(b)}
\resizebox*{1.0\columnwidth}{!}{\includegraphics{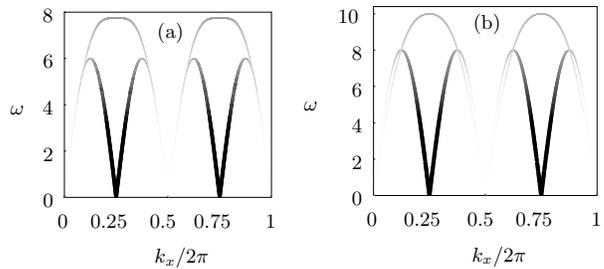}}
\caption{DS2:  Diagonal, site-centered stripes of spacing $p=2$.  
The plots are for twinned stripes,
summing contributions along $(k_x,k_x)$ and $(k_x,-k_x)$.
(a)$J_c=J_b$ and  $J_d=-0.5J_b$;
(b)$J_c=2J_b$ and  $J_d=-0.5J_b$. 
\label{fig:DS2}}
\end{figure}

Another key feature of
nonzero couplings $J_c$ and $J_d$ for site-centered stripes
is the symmetry of the spin wave structure,
as shown in Fig.~\ref{plot3d}.   In the limit where $J_c=J_d=0$,
the spin wave dispersion is symmetric under $90^o$ rotations,
as shown in Fig.~\ref{plot3d}(a).
However, when either $J_c$ or $J_d$ or both are nonzero, the symmetry
is broken, and the spin wave structure is now only symmetric under $180^o$ rotations,
as shown in Fig.~\ref{plot3d}(b).  
This means that for any nonzero $J_c$ or $J_d$,
the spin wave velocity of the acoustic mode dispersing out of $Q_{\rm AF}=(0.5,0.5)$
is different parallel and perpendicular to the stripe direction.  
In the presence of stripe twins, the two velocities will appear as two branches 
in plots of $\omega$ {\em vs.} $\vec{k}$, as shown in Fig.~\ref{fig:DS2}.

The magnon dispersion from Eq.~(\ref{model}) can  be solved analytically,
and for DS2 we find that 
\begin{equation}
\omega(k_x,k_y)=2\sqrt{A^2-B^2},
\end{equation}
where 
\begin{eqnarray}
A&=&2J_b+J_c-J_d+J_d\cos(k_x-k_y) \nonumber \\
B&=&J_b\cos(2k_x)+J_b\cos(2k_y)+J_c\cos(k_x+k_y).
\end{eqnarray}
There are two different spin wave velocities for the cones emanating
from the IC peak $Q_{\rm IC} = (0.25,0.25)$ and symmetry-related points,
one corresponding to spin wave velocities 
perpendicular to the direction of the domain walls ({\em i.e.} perpendicular
to the stripes), and the other parallel to the direction of the domain walls,
\begin{eqnarray}
v_{\perp}&=&4(2J_b+J_c) \nonumber \\
v_{\parallel}&=&4\sqrt{(2J_b+J_c)(2J_b-J_d)}~.   
\end{eqnarray}

\begin{figure}[htb]
\resizebox*{1\columnwidth}{!}{\includegraphics{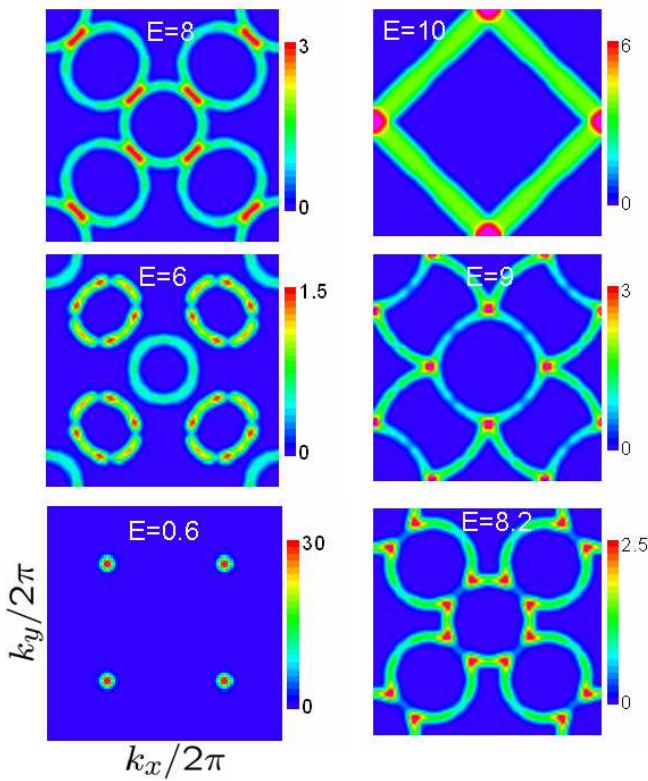}}
\caption{(Color online) Constant energy cuts with windows of $0.1J_bS$ for
twinned diagonal, site-centered stripes of spacing 
$p=2$ at $J_c=2 J_b$ and  $J_d=-0.5J_b$. 
The energy $E$ is in units of $J_b S$.
\label{cut.ds2}}
\end{figure}

In Fig.~\ref{fig:DS2}, we show the expected dispersions and 
scattering intensities for DS2.
Plots are shown for twinned stripes, summing the contributions 
parallel  and perpendicular to the stripe direction, {\em i.e.} along
the $(k_x,-k_x)$ and $(k_x,k_x)$ directions, respectively.  
Because $J_c$ and $J_d$ are nonzero, there is an apparent optical mode.  
Note that it is not a true optical mode, since 
in this configuration there are only two spins per unit cell, 
leading to only one (twofold degenerate) band.  
As in the bond-centered case, weight is forbidden at low energy at
the antiferromagnetic point $Q_{\rm AF}$, so that the spin-wave
cone emanating from this magnetic reciprocal lattice vector
has vanishing weight as $\omega \rightarrow 0$.  
In Fig.~\ref{fig:DS2}(a), we have set $J_c = J_b$ and $J_d = -0.5J_b$.
In Fig.~\ref{fig:DS2}(b), we use $J_c = 2J_b$ with $J_d = -0.5J_b$.
Notice that in panel (a) of the figure, 
the apparent optical mode is flat.

We have chosen the coupling constants with the following in mind:
The ``acoustic'' branch peaks at 
$\omega(3 \pi/4, 3 \pi/4) = 4 J_b + 2 J_c$.
The apparent ``optical'' mode peaks at 
$\omega(\pi/2, 3 \pi/2) = 4 \sqrt{(2J_b-J_d)(J_c-J_d)}$.
The data indicate that the apparent optical mode is higher in energy than
the top of the  ``acoustic'' part:
$\omega(\pi/2, 3 \pi/2) > \omega(3 \pi/4, 3 \pi/4)$, which implies that 
\begin{equation}
J_d \le (1/2)(2 + J_c - \sqrt{2}\sqrt{4 + J_c^2})
\label{eqn:constraint1}
\end{equation}
when $J_b = 1$.  
However, the extra mode above $50$meV is not too high in energy,
so parameters need to be chosen so as to satisfy this constraint,
but remain close to the equality.  

We also require the  apparent ``optical'' branch
to be concave, since there is no evidence of a dip in the extra mode.  
This requirement gives 
\begin{equation}
{\partial^2 \over \partial k_x^2} \omega (k_x, -k_x) = 
{4 (J_d - 2)(J_c - 2 -2J_d) \over \sqrt{-(J_c - J_d)(J_d -2)}}\le0~,
\end{equation}
resulting in the constraint that
\begin{equation}
J_d \le {1 \over 2}J_c - 1~.
\label{eqn:constraint2}
\end{equation}

\begin{figure}[t]
\psfrag{w}{$\omega$}
\psfrag{kx}{$k_x/2\pi$}
\psfrag{0.25}{$0.25$}
\psfrag{0.5}{$0.5$}
\psfrag{0.75}{$0.75$}
\psfrag{0}{$0$}
\psfrag{1}{$1$}
\psfrag{2}{$2$}
\psfrag{3}{$3$}
\psfrag{4}{$4$}
\psfrag{6}{$6$}
\psfrag{8}{$8$}
\psfrag{10}{$10$}
\psfrag{a}{(a)}
\psfrag{b}{(b)}
\resizebox*{1.0\columnwidth}{!}{\includegraphics{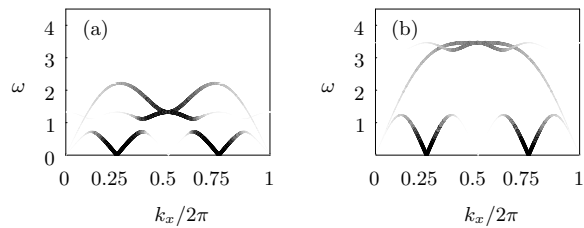}}
\caption{DB2:  Diagonal, bond-centered stripes of spacing $p=2$.
(See Fig.~\ref{lattice}(b).)  The plots are for twinned stripes,
summing contributions along $(k_x,k_x)$ and $(k_x,-k_x)$.
(a) Very weak coupling across the domain wall, with $J_b = -0.1 J_a$.
(b) Weak coupling across the domain wall, with $J_b = -0.5J_a$.  
\label{fig:DB2}}
\end{figure}

As long as the second constraint, Eqn.~\ref{eqn:constraint2},
is satisfied in the range $0\le J_c \le 2 J_b$, then the first constraint, 
Eqn.~\ref{eqn:constraint1},
is also satisfied.  
To describe the data, then, we find that we need a sizeable $J_c$, on the order of 
$J_b$.
This makes $J_c$ significantly larger than that reported at the lower doping $x=1/3$,
where diagonal site-centered stripes of spacing $p=3$ (DS3) were used
to explain the data successfully.\cite{boothroyd05} 
Although the fits in Ref.~\onlinecite{boothroyd05} were 
good for $J_c =0$ and were not significantly improved
by letting $J_c$ increase to $J_c \approx 0.5 J_b$, the data were still well described
using a nonzero $J_c$.
We find that the data at doping  $x=1/3$ are also well described by taking
$J_c$ to be as large as $J_b$ or even $2 J_b$
as in our Fig.~\ref{fig:DS2}.

We also find that we need $J_d<0$, {\em i.e.} the diagonal coupling
parallel to the stripes needs to be ferromagnetic, in order to 
describe the data.  If one considers the diagonal spin couplings $J_c$ and $J_d$ to be derived from, {\em e.g.},
a perturbative treatment of a single-band Hubbard or three-band Emery model on a square Ni-O lattice,
we expect $J_c = J_d$.  Given that the spin ground state breaks the $4$-fold rotational symmetry
of the square lattice to only $2$-fold rotational symmetry, any finite spin-lattice coupling
results in the two diagonal directions being inequivalent, and leads to $J_c \ne J_d$.  
For weak rotational symmetry breaking of the square lattice in the perturbative regime,
one expects $J_c = J_o + \epsilon$ with $J_d = J_o - \epsilon$ where $\epsilon$ is small compared to $J_o$, 
so that the anisotropy between the two diagonal coupling directions is weak as well.  
(This preserves the spin ground state of Fig.~\ref{lattice}(a).)
We find, however, that this regime of the coupling constants leads to the apparent optical mode
being too low in energy to capture the data.  
This may indicate that the materials are far from the perturbative limit of the 
single-band Hubbard or three-band Emery model.


Fig.~\ref{cut.ds2} shows  constant energy cuts for
DS2 corresponding to the parameters in Fig.~\ref{fig:DS2}(b).
An important feature of this configuration is that 
although the antiferromagnetic point $Q_{\rm AF} = (0.5,0.5)$ is a magnetic reciprocal lattice vector,
zero-frequency weight is forbidden there by symmetry,
since the stripes have no net N\'eel vector at $Q_{\rm AF}$.  
Combined with the fact that there is no optical band,
the DS2 configuration {\em can never have spectral weight at
the antiferromagnetic point} $Q_{\rm AF}$, even at finite frequency.
Notice that as energy is increased in Fig.~\ref{cut.ds2}, a 
faint spin wave cone emerges from $Q_{\rm AF}$ in a 
light ring of scattering, but none of the constant energy
plots show any weight at $Q_{\rm AF}$.  
This is consistent with the constant energy plots 
for La$_{3/2}$Sr$_{1/2}$ NiO$_4$
shown in Ref.~\onlinecite{freeman05a}.
By contrast, the corresponding bond-centered configuration
(DB2) shown in Fig.~\ref{lattice}(b)  has an optical band which
displays a saddlepoint at $Q_{\rm AF}$, 
and rather large scattering intensity at finite frequency at $Q_{\rm AF}$ as 
a result.  (See Fig.~\ref{fig:DB2} of this paper, as well as Fig.~8 of our previous paper,
Ref.~\onlinecite{yaocarlson06b}.)

While diagonal, site-centered stripes of spacing $p=2$ (DS2) are able to
account for the behavior of the apparent optical mode observed to disperse away from
$Q_{\rm AF}$ in Fig.~3 of Freeman {\em et al.}\cite{freeman05a}, 
there are two other high energy features which this model has not captured.
One is the asymmetry in intensity observed above $30$meV
in the spin wave cones emanating from the main IC peaks. 
The other is a mode in the $31-39$meV range propagating away from
$(h,k)$ structural reciprocal lattice points. 
These (as well as the apparent optical mode) have been attributed to discommensurations in the magnetic order.\cite{freeman05a}

In Fig.~\ref{fig:DB2}, we show the expected  dispersions and intensities
for the diagonal, {\em bond-centered} stripes (DB2) shown in Fig.~\ref{lattice}(b).  
This configuration has a true optical mode.  
Fig.~\ref{fig:DB2}(a) shows weak coupling across the charged domain walls,
with $J_b = -0.1 J_a$, and Fig.~\ref{fig:DB2}(b) has somewhat stronger coupling across
the domain walls, with $J_b = -0.5 J_a$.  Note that in the bond-centered case,
couplings across the domain walls are ferromagnetic.  
Results are shown for twinned stripe patterns, summing the contribution 
parallel  and perpendicular to the stripe direction, {\em i.e.} along
the $(k_x,-k_x)$ and $(k_x,k_x)$ directions, respectively.  
Although there is a reciprocal lattice vector at $Q_{\rm AF}$,
weight is forbidden there at zero frequency, since the N\'eel vector
switches sign across the domain walls.    
We have reported the 
analytic form of the spin wave dispersion in this case
in a previous publication.\cite{erica04}

Because this spin configuration
is only $180^o$ symmetric, the spin-wave velocity emanating from $Q_{\rm AF}$
is different parallel and perpendicular to the stripe direction.  
However, the branch emanating from $Q_{\rm AF}$ in the direction parallel to stripes
has so little weight as to be effectively invisible in the plots. 
This configuration displays a true optical mode because 
there are four spins in the unit cell. The optical mode has a {\em saddlepoint} at
$Q_{\rm AF}$ and finite energy, with increased weight at the saddlepoint. 
For weak coupling across the domain walls ($|J_b|<|J_a|$), the
optical mode always displays significant weight at $Q_{\rm AF}$ at
finite frequency.  This is not supported by the data\cite{freeman05a},
which at the energies measured
display no scattering at $Q_{\rm AF}$ and finite frequency.  
This likely indicates that the domain walls are not bond-centered in this material,
but are probably site-centered.


In conclusion, we have used linear spin-wave theory to 
describe the magnetic excitations recently observed
in neutron scattering\cite{freeman05a}  on 
La$_{3/2}$Sr$_{1/2}$ NiO$_4$.
Many features of the data were captured in the spin wave analysis of Ref.~\onlinecite{freeman05a}.
Other features, including an apparent optical mode dispersing away
from $Q_{\rm AF}$ above $50$meV, 
were attributed to discommensurations in the spin order.  
We show here that the apparent optical mode 
may also be captured in linear spin wave theory by using a spin coupling
configuration that preserves the symmetry of the spin ground state.
Namely, we have shown that diagonal, site-centered stripes of
spacing $p=2$ capture this mode when the pattern of couplings is $2$-fold symmetric.
This is because the $2$-fold symmetric coupling pattern gives rise to 
two different spin-wave velocities ({\em i.e.} $v_{\parallel} \ne v_{\perp}$)
emanating from the antiferromagnetic point $Q_{\rm AF}=(0.5,0.5)$.  
For twinned samples, the two velocities are simultaneously visible,
and the higher velocity mode $v_{\parallel}$ parallel to the stripes is responsible for the ``extra'' scattering
above $50$meV.  
Furthermore, this configuration is forbidden to display scattering at
the antiferromagnetic point $Q_{\rm AF}$, whereas
bond-centered stripes have a true optical mode with significant weight at $Q_{\rm AF}$,
which is not supported by the data.  We therefore conclude that the magnetic excitations observed
in Ref.~\onlinecite{freeman05a} are consistent with site-centered stripes,
but not with bond-centered stripes..  

It is a  pleasure to thank  D.~K.~Campbell and 
A.~T.~Boothroyd for helpful discussions. 
This work was supported by Boston University (DXY), and by the Purdue Research Foundation (EWC). 
EWC is a Cottrell Scholar of Research Corporation. 

\bibliographystyle{forprb}

\end{document}